\documentclass[12pt,a4paper]{article}
\usepackage{amsmath, graphicx, natbib, float, subcaption, color, afterpage, lineno, enumerate, multirow, adjustbox, xcolor, setspace, authblk, ulem}
\usepackage[toc,page,titletoc]{appendix}
\usepackage[labelfont=bf]{caption}
\usepackage{chngcntr}
\usepackage[margin=1in]{geometry}
\usepackage{orcidlink}
\usepackage{booktabs} 

\newenvironment{keyword}
    {\noindent\textbf{Keywords: }\ignorespaces}
    {\par}

\begin{document}

\title{RNN-DAS: A New Deep Learning Approach for Detection and Real-Time Monitoring of Volcano-Tectonic Events Using Distributed Acoustic Sensing} 

\author[1]{Javier Fernández-Carabantes
\orcidlink{0009-0000-1449-4361}}
\author[1]{Manuel Titos
\orcidlink{0000-0002-8279-2341}}
\author[2]{Luca D'Auria
\orcidlink{0000-0002-7664-2216}}
\author[1]{Jesús García
\orcidlink{0009-0009-9992-1494}}
\author[1]{Luz García
\orcidlink{0000-0001-5904-5412}}
\author[1]{Carmen Benítez
\orcidlink{0000-0002-5407-8335}}

\affil[1]{\small Department of Signal Theory, Telematic and Communications,
University of Granada, 18010 Granada, Spain}
\affil[2]{\small Instituto Volcanológico de Canarias (INVOLCAN), 38320, Puerto de la Cruz, Tenerife, Canary Islands, Spain}

\date{}

\maketitle

\begin{abstract}
In this article, we present a novel Deep Learning model based on Recurrent Neural Networks (RNNs) with Long Short-Term Memory (LSTM) cells, designed as a real-time Volcano-seismic Signal Recognition (VSR) system for Distributed Acoustic Sensing (DAS) measurements.
The model was trained on an extensive database of Volcano-Tectonic (VT) events derived from the co-eruptive seismicity of the 2021 La Palma eruption, recorded by a High-fidelity submarine Distributed Acoustic Sensing array (HDAS) near the eruption site. The features used for supervised model training, based on signal energy average in frequency bands, effectively enable the model to leverage spatial contextual information and the temporal evolution of volcano-seismic signals provided by the DAS technique.
The proposed model not only detects the presence of VT events but also analyzes their temporal evolution, selecting and classifying their complete waveforms with an accuracy of approximately 97\% for correctly detected and classified VT events.
Furthermore, the model has demonstrated robust performance in generalizing to other time intervals and volcanoes, enabling continuous real-time monitoring of seismicity. Such results highlight the potential of using RNN-based approaches with LSTM cells for application to other active volcanoes, enabling fast, automatic analysis with low computational requirements and the need of minimal retraining, for the creation of labeled seismic catalogs directly from DAS measurements. This represents a significant advancement in the use of DAS technology as a viable tool to study active volcanoes and their seismic activity.

\end{abstract}


\newpage

\section{Introduction}

The application of seismology to the study of volcanic phenomena has proven to be one of the most effective tools for obtaining valuable geophysical information about the dynamic behavior of volcanoes and understanding their eruptions \citep{iyer1992}. This success is attributed to the diverse origins of seismic signals generated by magma transport, changes in hydrothermal regimes, and pressure variations within the volcanic edifice \citep{chouet2003, chouet2013}.  
Among the different types of events, those known as Volcano-Tectonic (VT) events are the most similar to traditional tectonic earthquakes, sharing the same source mechanism and exhibiting P and S phases. These events are characterized by high frequencies and are of significant interest for monitoring volcanic activity \citep{white2016, roman2016}. VT events are caused by the fracturing of volcanic materials or the rupturing of nearby faults in response to dynamic changes within the volcanic system. Consequently, they can serve as precursors to a volcanic eruption in the form of seismic swarms, highlighting the importance of their study. 
However, their recordings differ from classical tectonic events due to the high heterogeneity of the volcanic medium through which they propagate \citep{zobin2003, lahr1994, almendros2007, afnimar2022}.

Detection and localization of volcano-seismic events have traditionally required a large spatial deployment of broadband seismic stations in the volcanic environment \citep{dreier1994, chouet1997}, either as dense networks or seismic arrays \citep{ibanez1997, almendros2002}. However, such configurations are expensive and challenging to maintain over the long term, being influenced by meteorological noise or ground tilt. As a result, most volcanoes rely on limited permanent station deployments, which may be insufficient to fully monitor volcanic unrest \citep{wassermann2022, debarros2013}.

In this context, promising new methodologies have been developed in recent years within the field of acoustics to address the aforementioned deployment challenges. Of particular interest is the Distributed Acoustic Sensing (DAS) technique \citep{zhan2020}. By measuring deformations along an optical fiber cable, using laser pulse emissions through an interrogator and analyzing Rayleigh scattering, produced at imperfections in the fiber caused by wavefront propagation \citep{hartog2017, lindsey2020}, it is possible to perform seismic event measurements in an array-like configuration of a single channel with unprecedented spatial and temporal coverage \citep{fichtner2023, williams2019}. This approach enables a logistically viable deployment across diverse environments, including glaciers \citep{walter2020}, the seafloor \citep{lior2021, nakano2024}, and highly heterogeneous and topographically complex settings such as volcanic terrains \citep{klaasen2021, biagioli2023}.
Furthermore, DAS provides broadband response capabilities \citep{lindsey2020}, with strain responses comparable to those of traditional point-like stations in volcanic environments \citep{currenti2021}, high temporal sampling frequencies (up to kHz), spatial resolutions on the order of meters, and measurement distances on the order of kilometers with only one interrogator \citep{lindsey2017}. These attributes explain the growing interest in applying DAS to volcanic seismology, with promising results demonstrated in locating tremor and explosion events using array techniques at Stromboli \citep{biagioli2023}, hypocenter localization and site response analysis at Azuma volcano \citep{nishimura2021}, very long period events at Vulcano \citep{currenti2023}, degassing-related events \citep{jousset2022}, submarine volcanic-related observations \citep{nakano2024} or structural imaging \citep{jousset2018}.

One of the most complex objectives in volcanic seismology is the development of robust databases of well-classified events using recognition pattern mechanisms to characterize them. Traditionally, the methodology employed for event detection and classification involved manual picking performed by expert geophysicists with years of experience \citep{ibanez2000}. This approach, while effective, is extremely time-consuming and further complicated by the vast amount of data generated by the high spatiotemporal sampling of DAS. In seismo-volcanic applications, this poses significant challenges for analyzing the recorded seismicity, both in terms of human and computational resources.  
This challenge marked the initial point where the use of Machine Learning techniques became important and proved to be game-changers. Their use has expanded and demonstrated efficacy in traditional volcanic seismology for detecting and classifying various types of events through different models: Support Vector Machines \citep{masotti2006, lara-cueva2017}, Decision Trees \citep{lara-cueva2016}, and Hidden Markov Models \citep{ibanez2009, beyreuther2008}.  
More recently, neural network models have facilitated the application of Deep Learning techniques to the classification of seismo-volcanic events, yielding promising results \citep{malfante2018, esposito2018, titos2018a}. Convolutional Neural Networks (CNNs), based on filter operations to extract signal features, have also been successfully explored \citep{Titos2019, Canario2020, Lara2021}. Additionally, Recurrent Neural Networks (RNNs), capable of leveraging the temporal evolution of seismo-volcanic signals for classification, have shown successful implementation \citep{titos2018b}.

Despite the success of Deep Learning in detecting and classifying volcanic events from traditional seismic recordings, the same cannot be said for its application to DAS. To date, no specific models exist for detecting and classifying VT events in DAS near volcanoes. This limitation is attributed to the novelty of the technique and the complexity of managing the large volumes of data involved, rather than a lack of interest \citep{biondi2024}.  
However, there are models adapted from traditional seismology to DAS that have shown promising results in event detection. Notable examples include the CNN-based model PhaseNet-DAS \citep{zhu2023} and a transformer-based model \citep{corsaro2024}, which are focused on automatic picking.  
This gap has led some researchers to rely on manual picking \citep{nishimura2021} or well-established methods in traditional seismology, such as the STA/LTA approach \citep{allen1978, biagioli2023, jousset2022, klaasen2021, nakano2024}, when studying volcanic related seismicity with DAS. While these methods can be effective, they require carefully tuned parameters \citep{trnkoczy2009} and are negatively affected by low Signal-to-Noise Ratios (SNR). A common issue with DAS as data tend to be noisier than traditional seismic measurements, with additional challenges posed by instrumental noise and signal attenuation due to the epicentral distance of the event to the DAS array \citep{baba2024, zhu2023}.

For this reason, in this article, we present a Deep Learning model (RNN-DAS) based on recurrent neural networks with Long Short-Term Memory (LSTM) cells \citep{lecun2015}. The model was trained in a supervised fashion using an automated database of events detected by an underwater high-fidelity DAS array deployed 10 km appart from the site of the destructive Tajogaite volcano during the eruption in 2021 on the island of La Palma, Spain \citep{troll2024, carracedo2022}. This model has been specifically designed to function as a real time Volcano-seismic Signal Recognition (VSR) system. It leverages features that allow the network to utilize the high spatial and temporal information provided by DAS without requiring adjustments for site effects or low SNR. Unlike previous models, the proposed one not only detects events but is also capable of selecting the waveform of the event, providing precise identification and extraction of the signals of interest. The methodology adapts the approach proposed by \cite{titos2018b} to enable the detection, classification, and real-time monitoring of events based on signal energy within frequency bands. This system has demonstrated excellent performance, achieving results comparable to those of visual detection by experts using conventional seismic stations.  

As an increasing number of people are exposed to the dangers of volcanic eruptions \citep{small2001}, the relevance of methodologies like this system grows significantly. These systems enable the automatic analysis and creation of volcano-seismic signal databases recorded by early warning systems in near-real time. Such capabilities allow geophysicists to evaluate the evolution of volcanic activity, facilitating the implementation of civil protection plans to mitigate material damage and save human lives \citep{brown2017, berz2001, hewitt2014, lara-cueva2016}.

The rest of this paper is structured as follows. Section 2 provides a brief explanation of the fundamentals of the DAS technique. Section 3 focuses on the capability of deep learning architectures to monitor continuous signals, justifying the use of RNNs with LSTM units to address the challenge of modeling the temporal evolution of seismic signals recorded by DAS. Section 4 describes the experiment conducted, including the characteristics of the HDAS used, the methodology for recording and selecting seismic events, the approach for extracting relevant features, the automatic labeling process for supervised model training and the generalization tests implemented to study the model capabilities. Section 5 presents the results and discussions of the model performance. Finally, Section 6 concludes the study.

\section{Distributed Acoustic Sensing (DAS) Technique}

In this section, the fundamentals of the Distributed Acoustic Sensing (DAS) technique are briefly presented. For a  detailed explanation of its methodology and the underlying physical-mathematical principles, readers are encouraged to consult the works by \cite{lindsey2020, zhan2020, shang2022}.

DAS is a novel technology recently applied in seismology, enabling continuous monitoring of strain ($\varepsilon$) or strain rate ($d\varepsilon/dt$) along a fiber optic cable when an acoustic wavefront (such as the one produced by seismic elastic waves from an earthquake) passes through it over distances of tens of kilometers. DAS offers ultra-dense spatiotemporal sampling at a low cost and with efficient deployment. Apart from the optic fiber cable, the other key element is the DAS interrogator. Which is positioned at the origin of the fiber optic cable, and continuously emits pulses of laser light. When a wavefront encounters the fiber and induces perturbations, natural imperfections within the cable cause Rayleigh scattering of the light pulses. A portion of this scattered light is retro-dispersed back to the interrogator. By measuring the phase shift of these retro-dispersed pulses, the strain or strain rate is calculated using interferometric techniques, with the imperfections serving as strain sensors distributed along the fiber.

This process enables probing over distances of up to tens of kilometers along the fiber, with temporal sampling frequencies reaching the kHz range and spatial sampling at the meter level. As a result, the DAS measurement can be interpreted as that of a strainmeter array, offering broadband characteristics, high fidelity of waveforms, and a good strain transfer response from the ground to the cable, making it suitable for seismological studies.  
The measurement with a DAS array depends on various instrumental and deployment parameters. While different models exist depending on the manufacturer, in most cases, a single DAS interrogator can probe distances of up to 50 km. The characteristics of the emitted laser pulses, such as their emission rate, width, and power, are also crucial. DAS provides distributed measurements, meaning that the spatial points or DAS channels where strain or strain rate is measured are located at the midpoint of what is known as the Gauge length ($L_G$), which is the spatial interval over which strain is computed (typically on the order of tens of meters). The Gauge length is related to the width of the laser pulse, and the distance between DAS channels, defined by the step between Gauge lengths, determines the spatial sampling interval ($dx$). Another relevant factor is the temporal sampling frequency ($dt$) and the time over which strain is derived along the channels, known as derivation time, used for computing the strain rate.  

In addition to these advantages, the use of DAS interrogators on pre-existing optical fibers, commonly referred to as \textit{dark fibers}, offers an easy and cost-effective application method. However, there are certain limitations inherent to the technique that present challenges. First, if the full potential of the technique is exploited, the volume of data generated can be immense, reaching terabytes per day. Further challenges include potential site effects, coupling effects with the ground, meteorological drifts in the measurements, and the precise location of the DAS channels or fiber cable.

\section{Deep Architectures for Continuous Monitoring}

Deep learning methods, originating from the field of artificial neural networks (ANNs), offer significant potential for addressing complex challenges in continuous signal processing. Among the most prominent models are Recurrent Neural Networks (RNNs), Transformers, Convolutional Neural Networks (CNNs), and Time Convolutional Networks (TCNs).

While Transformers have demonstrated promise in modeling temporal dependencies, they lack the components necessary to fully capture local structures within sequences, leading to the development of hybrid approaches that combine RNNs with Transformers \citep{wang2019}. Similarly, CNNs, which are well-suited for analyzing spatial data, have also proven effective in recognizing seismic signals \citep{Titos2019, dokht2019}. However, limited by their input size for continuous monitoring, efforts have been made to integrate them with RNNs to enhance temporal signal recognition \citep{Canario2020}. TCNs, combining deep architectures and dilated layers, are capable of modeling temporal sequences; however, their use for long time series remains constrained by kernel size limitations \citep{lea2016, titos2024}.

RNNs, especially with advancements such as Long Short-Term Memory (LSTM) networks, continue to outperform Feedforward Neural Networks (FNNs) in tasks involving temporal dependencies. While RNNs represent an earlier type of architecture, they remain highly effective for many applications, particularly those requiring the modeling of sequential data, such as detecting seismic signals in continuous data streams \citep{lecun2015, schmidhuber2015}. Unlike traditional feedforward models, RNNs generate outputs not only from the current input but also from prior inputs, thanks to the hidden state that is updated over time and fed back into the network. This allows RNNs to process and store memories of arbitrary input sequences, making them suitable for tasks with complex temporal dependencies \citep{schmidhuber2015}. Additionally, RNNs have several advantages over more recent models, including lower hardware requirements for computation, greater explainability in neural activation patterns, and fewer parameters required compared to other architectures \citep{titos2024}. However, training RNNs can be challenging due to issues such as \textit{vanishing} or \textit{exploding} gradients, which limit their ability to model long-term dependencies \citep{pascanu2013}. To address these issues, LSTM networks were introduced, incorporating specialized gates that regulate the flow of information, enabling the network to effectively store, discard, or update information in the memory cell \citep{kamath2019}. This makes LSTMs an ideal choice for modeling temporal data with long-term dependencies. For a more in-depth explanation of the mathematical foundations of RNNs, readers are encouraged to consult appendix A.  

This ability to process sequential data has established RNNs as a state-of-the-art solution across various scientific fields, including speech recognition \citep{graves2013}, medical diagnostics \citep{choi2017}, meteorological forecasting \citep{balluff2020}, image generation \citep{gregor2015}, and genetics \citep{xu2007}. Furthermore, RNNs have proven effective in geophysical applications such as volcano monitoring, as demonstrated by \cite{titos2018b}. Given these strengths and their continued relevance despite the emergence of newer architectures, RNNs with LSTM remain an optimal choice for capturing the temporal evolution of seismic signals recorded by the DAS array, while maintaining high model explainability and simplicity.

\section{Experimental Design}

\subsection{Description of the Field Experiment: La Palma 2021 Eruption}
\begin{figure}[H]
    \centering
    \includegraphics[width=1\linewidth]{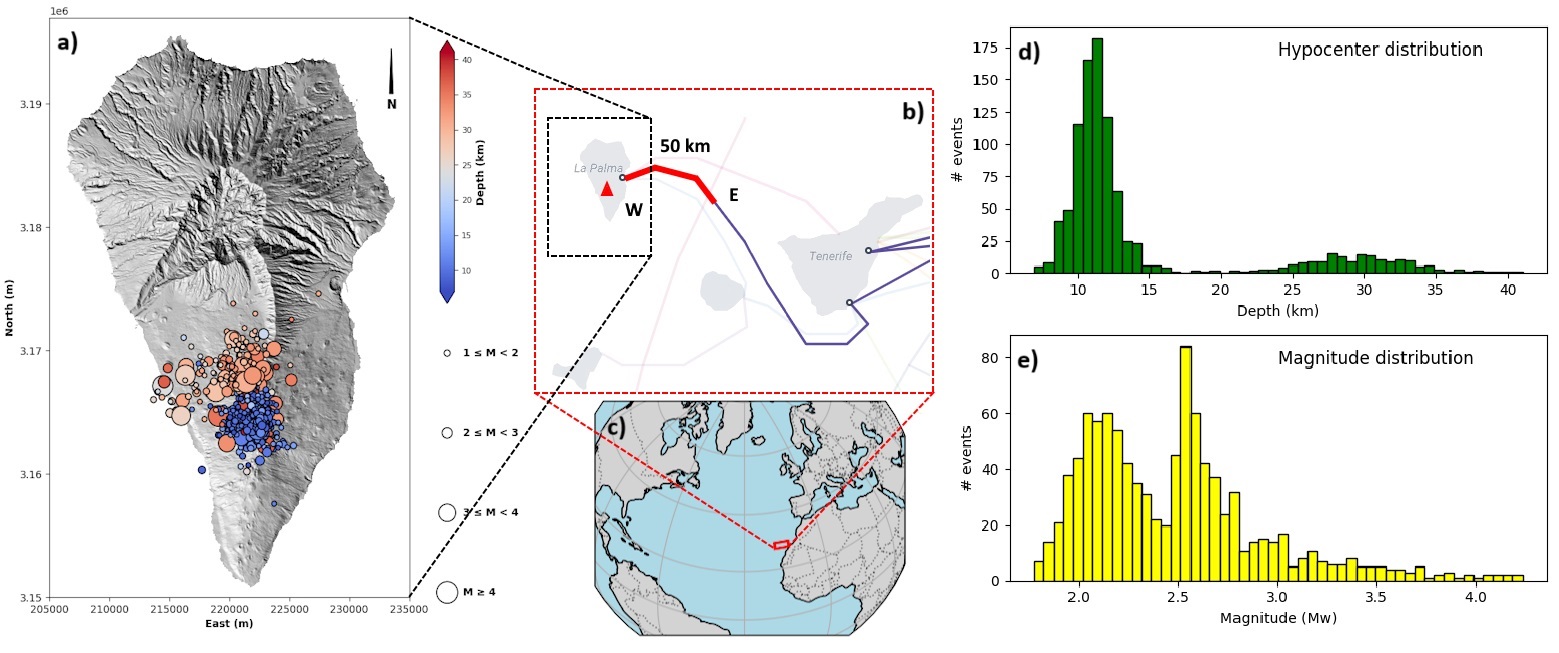}
    \caption{a) Map of La Palma showing the VT events used to train the model, where color indicates the depth of the hypocenter and the size of the epicentral circle represents the magnitude. The digital elevation model was obtained from \cite{copernicus2024}. b) Map of the portion of the submarine cable (red line) used for the HDAS array. Modified from \cite{barrancos2022}. c) Location map of the Canary Islands, Spain. d) Histogram of the hypocenter distribution for the events present in the model database used for RNN-DAS. e) Histogram of the magnitude distribution for the events present in the database used for RNN-DAS.
}
    \label{fig:map}
\end{figure}

The 2021 La Palma eruption occurred on the volcanic ridge of Cumbre Vieja, the most active area from a volcanological perspective of the  on the island of La Palma, Canary Islands, Spain. The eruption, which lasted from September 19 to December 13 2021, was characterized as the longest in the historical record of the island, the one with the highest amount of volcanic material ejected, and also the most destructive \citep{carracedo2022, delfresno2023}.

The eruption took place along a series of fissural openings with the emission of basaltic lavas, generating a cinder cone in the area of greatest effusive activity, known as the Tajogaite volcano \citep{carracedo2022}. The eruption was marked by different periods of explosivity \citep{romero2022} and was preceded by significant seismic activity. Since 2017, the seismic reactivation of the island had been evident, with swarms of VT events and uplift deformation indicating signs of major disruptions in the internal system of Cumbre Vieja \citep{torresgonzalez2020}. Finally, a week prior to the eruption, intense seismic activity of VT events occurred, associated with pressure changes in the internal systems of Cumbre Vieja due to magma ascent and migration toward the eruptive vents \citep{dauria2022}. Co-eruptive seismicity persisted in the form of different types of volcano-seismic events, including low-frequency volcanic tremor, long-period events, and VT events. The latter occurred in two depth clusters with a notable gap between them (see Figure \ref{fig:map} a)), driven by aseismic magma transport and consequent dynamical changes between shallow and deep reservoirs \citep{delfresno2023}.

During the eruption, the INstituto VOLcanológico de CANarias (INVOLCAN) and its partners installed, on October 19, a High-fidelity DAS (HDAS) system using a submarine fiber-optic cable that connects the islands of La Palma and Tenerife (see Figure \ref{fig:map} b) and c)). This HDAS, installed approximately 10 km from the Tajogaite volcano, recorded all co-eruptive seismic activity, including tremor, VT events, regional tectonic events, and teleseismic events. After the eruption, it continued recording measurements and remains operational to this day. The HDAS recorded data at a sampling frequency of 100 Hz, with a spatial sampling interval of 10 m, and an initial total distance of 30 km, later extended to 50 km via Raman amplification \citep{barrancos2022}.

\subsection{Database Description: Pre-Processing, Features and Labels Extraction Pipelines}

Using the seismic catalog of events recorded during and after the eruption by INVOLCAN, a total of 1,000 VT events recorded by the HDAS during November 2021 were selected to train the RNN-DAS model. These events exhibit a bimodal depth distribution (Del Fresno et al. 2023) and a magnitude distribution characterized predominantly by small to moderate magnitudes (see Figure \ref{fig:map} d) and e)). Several key aspects must be considered when training the model:
\begin{itemize}
    \item The presence of natural, anthropogenic, or internal noise can significantly affect the recordings of the events, particularly in HDAS array data, which are inherently influenced by low SNR conditions. This leads to a degradation of seismic signal quality \citep{bormann2013}.
    \item Site effects in the measurements, as the seafloor conditions where the HDAS is deployed influence how the events are recorded and their coupling with the ocean floor \citep{lior2021}. 
    \item Seismic waves are affected during their propagation by the medium through a series of effects that alter their recording. These effects are amplified in highly heterogeneous volcanic environments due to their abrupt topography and lateral velocity discontinuities, leading to attenuation, scattering, diffraction, and path deviation of the seismic energy \citep{durek1996, schwarz2019, almendros2001}.
    \item The system's performance heavily depends on the accuracy of the model parameters estimated from the training data \citep{titos2018b}.
\end{itemize}

\begin{figure}[H]
    \centering
    \includegraphics[width=1\linewidth]{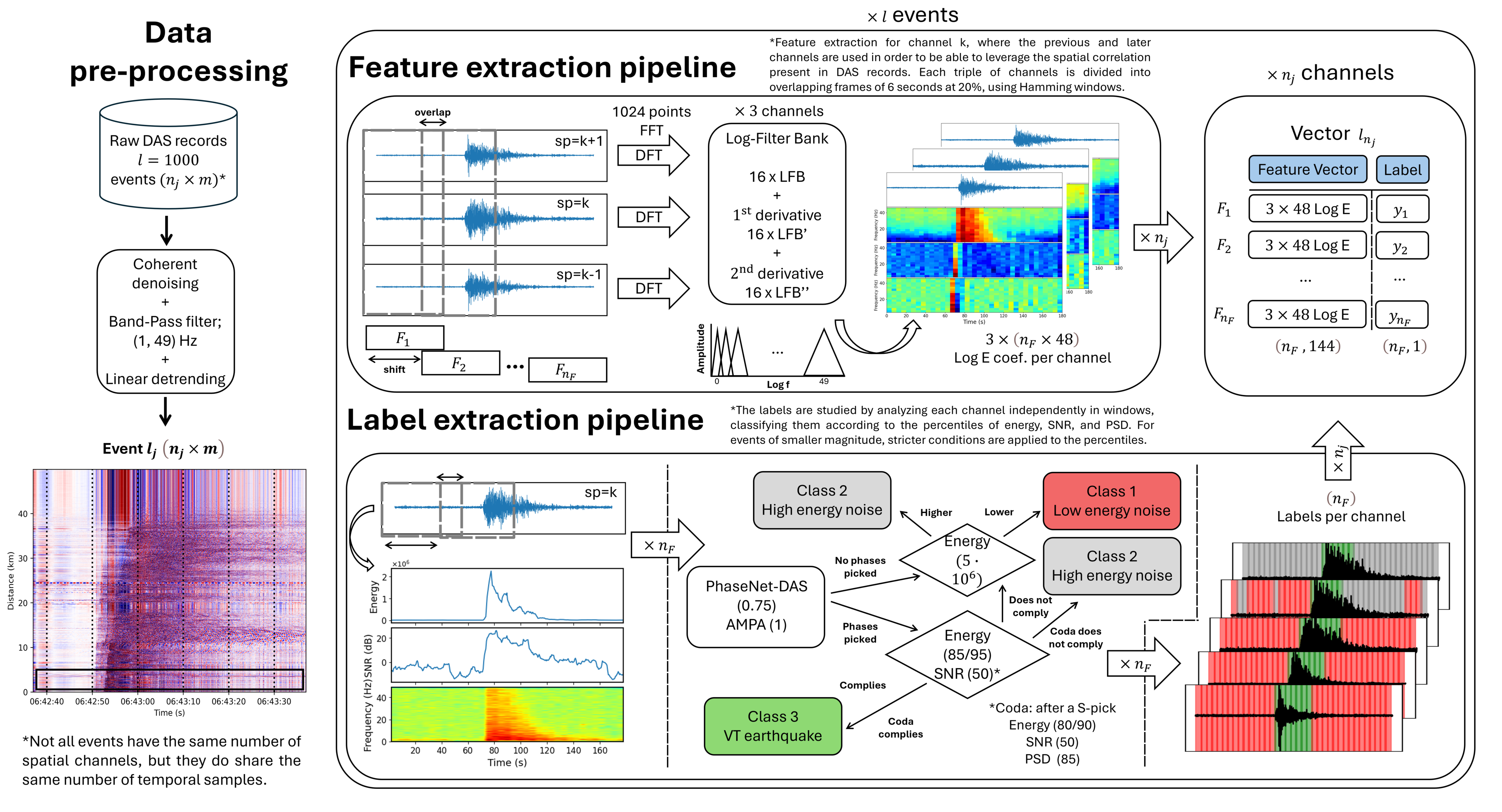}
    \caption{Schematic of the methodology employed in the preprocessing pipelines for DAS data, feature extraction representing energy coefficients that account for the temporal evolution and spatial coherence of the studied signals, and the label extraction pipeline for supervised training (with the threshold percentile values applied to each variable, depending on whether the event is of high or low magnitude). This procedure is followed for each of the $n_j$ channels of each event $l$ in the dataset, obtaining the feature and label values for each frame $F_k$ of the signals.}
    \label{fig:pipeline}
\end{figure}

The procedure for data pre-processing and the extraction of labels and features for model training is explained in detail below, being a modified version of the approaches proposed by \cite{titos2018b, benitez2006} (see Figure \ref{fig:pipeline}).

\begin{itemize}
    \item Data pre-processing pipeline. For each event, based on its origin time from the catalog, 3 minutes of HDAS signal are selected to ensure the complete signal is captured by including one minute before and one minute after the event. First, denoising is applied to remove coherent noise, followed by a double 4th-order causal band-pass filter between 1 and 49 Hz, as the focus is on detecting VT events. Finally, a linear detrending is applied to the data.

    \item Feature extraction pipeline. For the features to be learned by the RNN model, we use a signal parametrization based on Logarithmic Filter Banks (LFB), which allow the extraction of a coefficient vector estimating the energy of the signal in frequency bands with higher resolution at lower frequencies. This methodology has been proposed as relevant for the study of seismo-volcanic signals \citep{titos2018b, benitez2006, ibanez2009}, as opposed to directly learning features from the time domain of the signals, due to the nature of seismo-volcanic events, where temporal duration or volcanic system behavior can vary. Additionally, simultaneous recording of events may occur, making it complex to model temporal behavior directly from features extracted from the raw input data. For these reasons, signal parametrization remains a suitable approach.
    To obtain the feature vector of the signal associated with channel $k$, $\text{F}_k$, from the HDAS, we simultaneously use the records from channels $k-1$ and $k+1$. The signal is divided into $n_F$ frames using 6-second Hamming windows with 20$\%$ overlap. This configuration was chosen to strike a balance between the accuracy required for signal detection and the computational load of the model, enabling real-time analysis. For each of the $n_F$ frames, a 1024-point FFT is computed, producing 16 LFB coefficients, 16 coefficients for the first temporal derivative of the LFB vector, and another 16 for the second derivative, for each of the 3 channels. This results in a 144-component vector for each frame:
    \begin{align}
    & \text{F}_k^{n_F} = \big(\text{LFB}_i, \text{LFB}'_i, \text{LFB}''_i \ \big| \ i \in \{k-1, k, k+1\}\big) \Rightarrow \text{F}_k = (n_F, 144),
    \end{align}
    where in the second part of the equation, the features from all frames of channel $k$ are stacked into a matrix. Once this stacking is performed for all $n_j$ spatial channels of the HDAS, the result is a feature matrix for event $l$ with dimensions $\text{F}_l = (n_j, n_F, 144)$.
    This configuration was found to best model temporal sequences while leveraging the full potential of the spatial and temporal sampling offered by the DAS technique. By incorporating contiguous channels, coherent spatial information is utilized, and by analyzing the LFB coefficients along with their first and second derivatives, the evolution of energy, its velocity, and acceleration across the signal are studied. This has proven to be a feature set rich in information for the model. While using only the LFB coefficients yielded good results for high-SNR events, for lower-magnitude events, low-SNR signals, or highly attenuated signals, the inclusion of the derivatives significantly enhanced the model's ability to highlight energy changes caused by the arrival of seismic phases associated with a VT event, thereby improving the model's identification capability.

    \item Label extraction pipeline. Due to the large number of events included in the database and the waveform analysis required for each channel, the use of a manually labeled database for supervised training is unfeasible. Instead, the labeling process is performed automatically based on signal parameters. To achieve our objective, three distinct classes were defined for the classification problem: the first class corresponds to the presence of an event, while the remaining two refer to its absence. These latter classes are labeled as \textit{Noise Type 1} and \textit{Noise Type 2}, differentiated based on an energy threshold defined by observing differences between noisy and silent channels. This approach enables the distinction between low-energy natural seismic noise and high-energy noise potentially caused by external sources (e.g., anthropogenic activities) or by channels exhibiting strong site effects or poor coupling with the seafloor, thus ensuring a more robust signal analysis.

    As with the feature extraction process, each signal from channel $k$ is divided into $n_F$ frames of equal length via Hamming windowing, and a class label is assigned to each frame. Consequently, the label vector for each channel is represented as $L_k = (n_F, 1)$, and when stacking the labels for the $n_j$ channels in the DAS array, the resulting label matrix has a final dimension of $L_l = (n_j, n_F, 1)$.
    
    Considering that each selected signal sample contains a cataloged event, the class of each frame is determined based on its energy content, power spectral density (PSD), and SNR values relative to a set of thresholds. These thresholds are fine-tuned manually based on percentiles of the signal (see Figure~\ref{fig:pipeline}). 
    
    First, given the possibility that the event may be indistinguishable from background noise due to attenuation, the PhaseNet-DAS \citep{zhu2023} and AMPA models \citep{alvarez2013} are applied with thresholds of 0.75 and 1, respectively. These models are used to detect whether any P or S phases are present, or whether no event is detectable. If phases are identified, the frames exceeding the energy and SNR thresholds are labeled as VT, while the remaining frames are classified as noise. Based on their energy levels, the noise frames are further subdivided into the two defined noise classes. To enhance coda analysis, if an S-phase is detected, the thresholds are lowered, and an additional PSD condition with a minimum threshold is introduced to improve the detection of windows corresponding to the event's coda. It is worth noting that the thresholds for the metrics are adjusted based on event magnitudes. For events with magnitudes below 2.5, stricter energy thresholds are imposed by increasing the percentile threshold by 10\% to avoid mistakenly labeling frames without a signal.
    
    Energy estimation is performed by calculating the zero-lag autocorrelation within each frame, which corresponds to the squared sum of the signal coefficients \citep{dondurur2018}. The PSD is estimated frame-by-frame and normalized to a range between 0 and 1 using the minimum and maximum values of the studied time sample. The SNR is estimated in decibels (dB) for each frame, using a 3-minute noise sample from the HDAS as a reference.
\end{itemize}

\subsection{Design of the RNN-DAS Model}

To design the RNN models with LSTM cells, PyTorch \citep{NEURIPS2019_9015}, a Deep Learning framework implemented in Python, was used. Additionally, all database pre-processing prior to training, as described in Section~3.2, was also implemented in Python 3.7.  

The dataset used for training was split into training (64\%), validation (16\%), and test (20\%) sets using an 80-20 train-test split. Subsequently, the training set was further divided into training and validation subsets in an 80-20 ratio, with a random shuffle applied to balance the sets. 

The hyperparameters that yielded the best results were as follows: a batch size of 256, 256 hidden states, 1 layer depth for the model, and a variable learning rate \([0.1, 0.0001]\) controlled by a cosine scheduler. To compute training errors, softmax normalization was applied to predictions to obtain class probabilities, and the cross-entropy criterion \citep{hinton2012} was used. SGD was employed as the optimizer to adjust the model weights during each training epoch.  

Two different models were trained. The first one, referred to as RNN-DAS (1000), was trained for 100 epochs using the database of 1000 events described earlier, with the characteristics specified above. The second model, RNN-DAS (1150), involved fine-tuning the first model with a balanced database of 150 events, consisting of 50 events with magnitudes above 2.5, 50 events with magnitudes below 2.5 (both randomly selected from the original database), and 50 noise samples without cataloged seismic events. This fine-tuning was performed over 10 epochs, with the learning rate reduced by 10\% relative to the first model. The goal of this fine-tuning was to enable the RNN to also model continuous signals without events.  

An important consideration is ensuring the model’s applicability to DAS records with varying signal amplitudes and energy levels. To achieve this, the features of all events in the training set were equalized by computing the mean and standard deviation for each of the 144 features used by the model. These statistics were then applied to equalize the validation and test sets, as well as any input data for which the model made predictions.  

Given that each HDAS record consists of \(n_j\) channels and the event records for individual channels can differ, it was necessary to assign specific labels and features to each channel in the database. As a result, a significantly larger dataset of individual seismic waveforms was utilized for training, as outlined in Table~\ref{tab:events}. This process led to the creation of an automated benchmark database comprising over 3 million unique VT waveforms. This ensures robustness against potential outliers during feature and label extraction, resulting in models trained on a large variety of VT event waveforms.

\begin{table}[H]
    \begin{centering}
    \caption{Number of unique seismic samples used for model training.}
    \label{tab:events}
    \begin{tabular}{ccccc}
    \toprule 
    Model & Events & Training & Validation & Test\tabularnewline
    \midrule
    \midrule 
    RNN-DAS (1000) & 1000 & 2182459 & 545615 & 1169174\tabularnewline
    \midrule 
    RNN-DAS (1150) & +150 & 476160 & 119040 & 148800\tabularnewline
    \bottomrule
    \end{tabular}
    \par\end{centering}
\end{table}
Given the high computational load associated with training deep learning models and the large amount of DAS data involved in the training process, the task was carried out using an NVIDIA GeForce GTX 1080 Graphics Processing Unit (GPU), with CUDA version 9.0. The use of a less advanced GPU was intentional, aiming to demonstrate and develop a model capable of operating on most systems present in volcanic observatories worldwide or on less powerful devices, ensuring efficiency with minimal advanced hardware requirements.

To evaluate the performance of the model, a series of metrics \citep{kamath2019} are employed to assess the predictions. Specifically, the accuracy metric is used as the primary criterion to determine whether a new model is an improvement during training on the validation set, with the aim of avoiding overfitting. Additionally, the metrics of precision, recall, and F1-score are also studied to gain further insight into the model's behavior.

The used metrics are commonly derived from the confusion matrix. Based on its elements, these metrics are defined as follows:

\begin{itemize}
    \item Accuracy (ACC): Measures the proportion of correctly classified instances (both positive and negative) over the total number of predictions. It provides an overall sense of the model's correctness.
    \begin{equation}
    \text{ACC} = \frac{\text{True Positives} + \text{True Negatives}}{\text{Total Number of Events}}
    \end{equation}

    \item Precision (PR): Measures the proportion of correctly predicted positive instances out of all instances predicted as positive. It evaluates the quality of positive predictions.
    \begin{equation}
    \text{PR} = \frac{\text{True Positives}}{\text{True Positives} + \text{False Positives}}
    \end{equation}

    \item Recall (RC): Measures the proportion of correctly predicted positive instances out of all actual positive instances. It evaluates the model's ability to detect positive cases.
    \begin{equation}
    \text{RC} = \frac{\text{True Positives}}{\text{True Positives} + \text{False Negatives}}
    \end{equation}

    \item F1-score (F1): The harmonic mean of precision and recall, which provides a single metric that balances both concerns. It is especially useful when the class distribution is imbalanced.
    \begin{equation}
    \text{F1} = 2 \cdot \frac{\text{PR} \cdot \text{RC}}{\text{PR} + \text{RC}}
    \end{equation}
\end{itemize}

\subsection{Generalization tests}

To evaluate the adaptability and predictive capability of the model on data not used during training, two additional experiments have been designed. 

The first experiment examines data recorded by the same HDAS array during the Cumbre Vieja eruption, but from the month of October. This analysis focuses on continuous monitoring over a 24-hour period to ensure the robustness of the RNN-DAS model. The primary goal is to verify that, despite potential co-eruptive changes in the volcanic system and the resulting seismicity, the detection of events remains accurate. Event detections are compared with the existing seismic catalog derived from traditional seismic stations and the well-known PhaseNet-DAS model \citep{zhu2023}. Both models employ a probability threshold of $2/3$ to identify the onset of a phase or event, and a minimum interval of two RNN-DAS windows (9.6 s) between events is enforced. An event is considered valid if at least 500 DAS channels register a phase onset, following the methodology proposed by \citep{zhu2023}. 

The second experiment tests the RNN-DAS model at Stromboli volcano, utilizing a one-hour sample of records capturing explosive events and volcanic tremor \citep{IPGP.2023.lgl6bxby_2023}. Unlike the La Palma study, this DAS system is deployed on land. Explosions at Stromboli are frequent \citep{ripepe2008}, with energy spanning a broad frequency range (up to 10 Hz) \citep{neuberg1994}, and are caused by bursts of volcanic gases in the conduit \citep{blackburn1976}. These explosions release energy both acoustically and as seismic waves, the latter of which are recorded by the DAS array as a specific type of volcanic earthquake. Such earthquakes are typically spindle-shaped and are characterized by emergent onsets without clear S-wave arrivals \citep{biagioli2023}.
The objective of this experiment is to assess the generalization capabilities of the RNN-DAS model when applied to DAS measurements recorded on land and to evaluate its ability to detect seismo-volcanic events with similar energy content but originating from different source mechanisms. This investigation addresses the critical need for developing robust models that can generalize to other volcanic environments and adapt to varying seismic record types with minimal retraining. For this evaluation, the RNN-DAS model predictions were analyzed using a threshold of 0.5 for the predominant class to assess raw model performance. These results were directly compared with PhaseNet-DAS predictions. Considering that an event is detected following the same approach as the first experiment, but proportionally reducing the required number of simultaneously identified channels with events to 50 for the new DAS data.

\section{Experimental Results}
\subsection{Performance of the model}

First, the results of the metrics presented in Section 4.3 for both trained models are shown in Table \ref{tab:metricas}. These results refer to per-frame predictions and demonstrate good performance, with values exceeding 90\% in all cases. For both models, the RC and PR metrics exhibit lower performance, which can be attributed to the frames corresponding to the coda of the events.
The coda corresponds to the attenuated portion of the seismic signal, generated by scattering through the medium during its propagation, making it more similar to background seismic noise. This leads, in the case of RC, to errors caused by the insertion of false negatives when classifying that portion as noise. Similarly, the lower performance in PR can be explained by the introduction of false positives in the coda portion, where frames of seismic noise are confused with coda frames. This hypothesis is supported by the improved performance of all metrics in the RNN-DAS (1150) model, which, after being retrained using only seismic noise samples, is able to detect and differentiate coda frames from background noise with higher accuracy.
Despite these potential frame-level errors, the metrics' performance is high in both cases, demonstrating the ability of the RNN-LSTM architecture to model time series with spatial coherence recorded by the HDAS. The RNN-DAS (1150) model was selected as the better-performing model for the rest of the study.

\begin{figure}[H]
    \centering
    \includegraphics[width=1\linewidth]{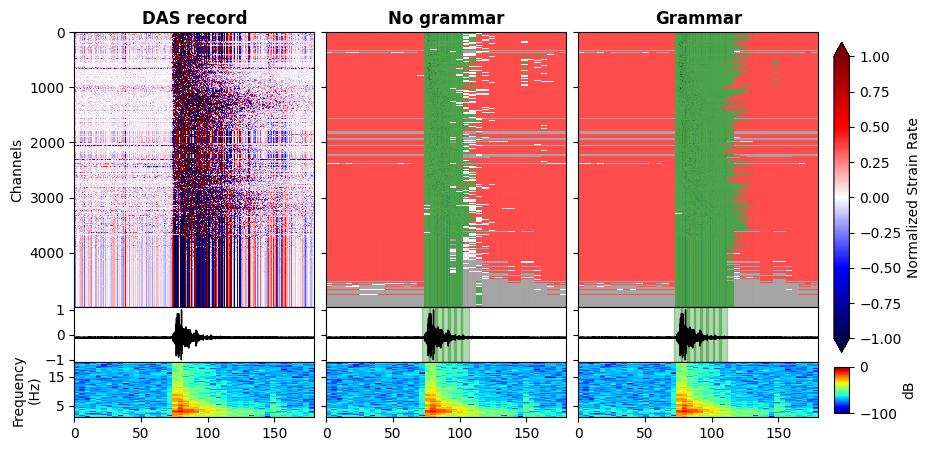}
    \caption{Example of the RNN-DAS model prediction with and without the grammar function applied. The original recording is 3 minutes long and corresponds to an event with $\text{M}_\text{l} = 4.04$. In the prediction outputs, the windows identified in green correspond to the VT class, those in red to noise type 1, and those in gray to noise type 2. If the 2/3 probability threshold for any class is not exceeded within a given window, the window is displayed as transparent. The normalized signal is shown alongside the straingram and spectrogram of channel 700, with the windows identified as VT for that channel highlighted in green.}
    \label{fig:grammar}
\end{figure}

The model outputs, for a DAS recording, a series of class-normalized probabilities for each channel and time frame. To further leverage the capabilities offered by the DAS technique for detection and to reduce potential prediction errors across channels or with the event's coda, a grammar function was developed to act on these class output probabilities (see Figure \ref{fig:grammar}). This function operates in two distinct ways. 
First, the method ensures consistency across channels by applying a filter to the probabilities of the VT or low-energy seismic noise classes for each time frame across all channels. If a threshold number of frames within a group of channels predict the same class, all frames that either predict a different class or have a predicted probability lower than a predefined threshold will be reassigned to the majority class within that group. This threshold value is manually set in advance to guide the reassignment process. Thereby ensuring robustness against individual frame prediction errors. Note that the second class of noise is not considered in this functionality, as it is specifically used to identify channels affected by high-amplitude noise components resulting from site effects or anthropogenic sources.
The second functionality is designed to improve coda prediction based on the previously discussed issues. Using a trigger-on and trigger-off approach with VT class probabilities across frames of a channel, probabilities are adjusted for frames that, after detecting a VT event with a channel-specific probability above the trigger-on threshold, are assigned to VT classes, even if the predominant class is seismic noise, as long as the VT class prediction probability is above the trigger-off threshold. This mechanism enhances the detection of the seismic coda.
This grammar function has proven to be a useful tool for improving the raw predictions of the model. For the La Palma case, it was found that frame probability thresholds of 2/3 for 50\% of the channels within groups of 10 channels (100 m of monitoring), with trigger-on and trigger-off thresholds for the coda of 0.9 and 0.05, respectively, provide satisfactory results.

\begin{table}[H]
\begin{centering}
\caption{Metrics per frame of learning models on the test set and total
training time.}
\label{tab:metricas}
\begin{tabular}{ccccccc}
\toprule 
Model & Loss & ACC ($\%$) & PR ($\%$) & RC ($\%$) & F1 ($\%$) & Time (s)\tabularnewline
\midrule
\midrule 
RNN-DAS (1000) & $0.1302$ & $95.94$ & $93.88$ & $90.63$ & $92.23$ & $81574$\tabularnewline
\midrule 
RNN-DAS (1150) & $0.0816$ & $97.46$ & $95.13$ & $92.86$ & $93.98$ & $1071$\tabularnewline
\bottomrule
\end{tabular}
\par\end{centering}
\end{table}

\begin{figure}[H]
    \centering
    \includegraphics[width=1\linewidth]{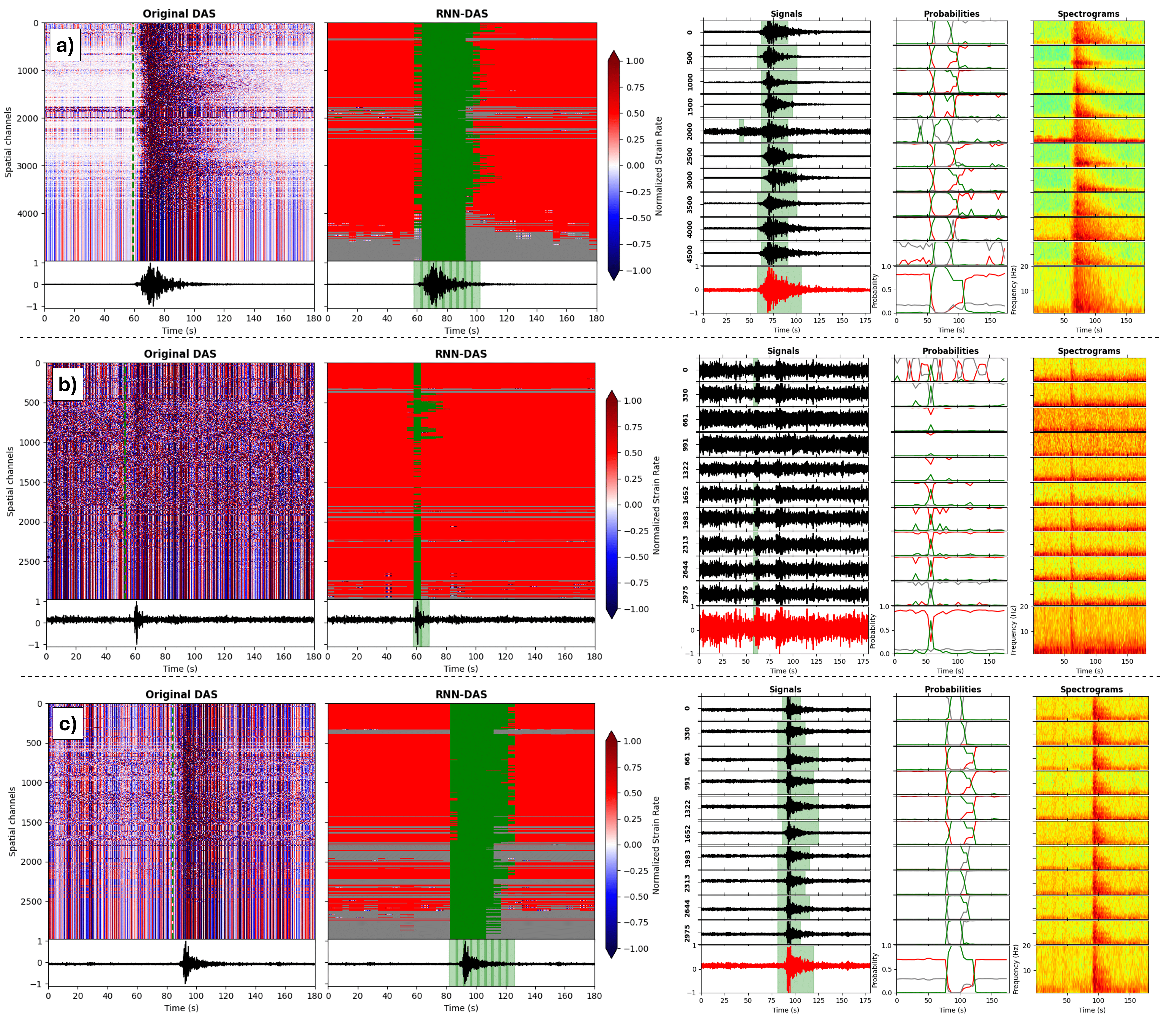}
    \caption{Examples of RNN-DAS model performance on different VT event recordings. Left: The original recording with a green dashed line marking the event origin time from the seismic catalog, alongside the prediction outputs for the three classes (green = VT, red = low-energy noise, gray = high-energy noise), showcasing the straingram from channel 1500. Right: Straingrams from various DAS channels with the regions identified as the VT class highlighted in green, the temporal evolution of the output probabilities per class provided by the model, and the spectrogram for each channel. The average signal is shown in red. a) Attenuated event with magnitude $\text{M}_\text{l} = 3.76$. b) Event recorded with low SNR and magnitude $\text{M}_\text{l} = 2.28$. c) Nearby event recorded with high SNR and magnitude $\text{M}_\text{l} = 2.75$.}
    \label{fig:model_performance}
\end{figure}

Figure \ref{fig:model_performance} illustrates examples of the model's performance on different types of VT event recordings from the month of November. The model demonstrates excellent detection capability in all cases and across the majority of DAS channels, with the use of the grammar function enhancing both coda detection and coherence.

It is worth noting that, despite the presence of high-amplitude seismic noise that can mask the signal, the model remains capable of accurately distinguishing the event. This is because the model predicts using the estimated energy based on frequency, allowing it to differentiate the event from the noise, even though its identification becomes more complex, as evidenced in case b).

\begin{figure}[H]
    \centering
    \includegraphics[width=1\linewidth]{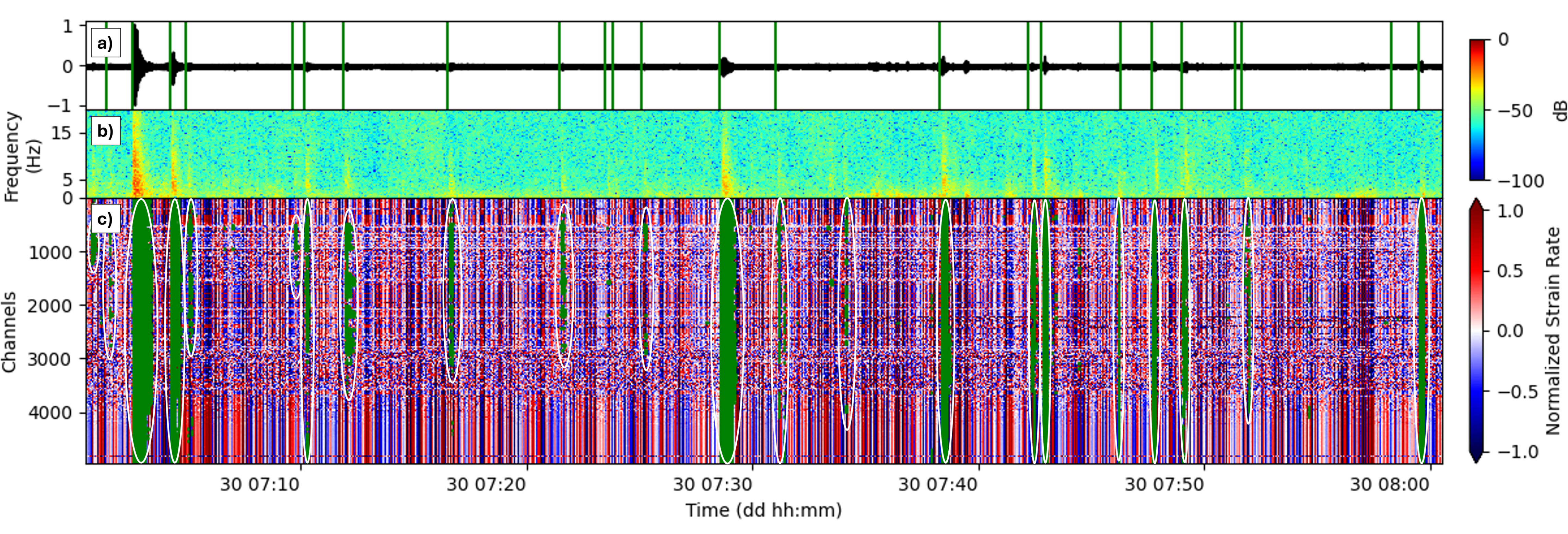}
    \caption{Example of the model's performance on a one-hour sample from October 30, between 07:00 and 08:00, featuring a main event of \(\text{M}_\text{l} = 3.22\). a) Straingram of channel 1500 with green lines indicating the origin time detected in the seismic catalog, along with its spectrogram in b). c) DAS signal with frames predicted as events shown in green, using a probability threshold of $2/3$, and events detected in at least 500 DAS channels highlighted with a white oval.}
    \label{fig:main_shock}
\end{figure}

Another case of interest when evaluating the model's performance is a sequence of events consisting of a main shock followed by lower-magnitude aftershocks. Figure \ref{fig:main_shock} presents such a sequence, where a primary event is followed by a large number of smaller events within a short time interval. It is observed that, despite the events occurring in close temporal proximity and exhibiting low SNR, the vast majority are correctly identified across most DAS channels. Only some events, which are significantly attenuated and detectable only by stations near the volcano, are not accurately detected by our model in the DAS records. Nevertheless, the results demonstrate a strong performance of the RNN-DAS model.

\subsection{Generalization capabilities}

\begin{figure}[H]
    \centering
    \includegraphics[width=\linewidth]{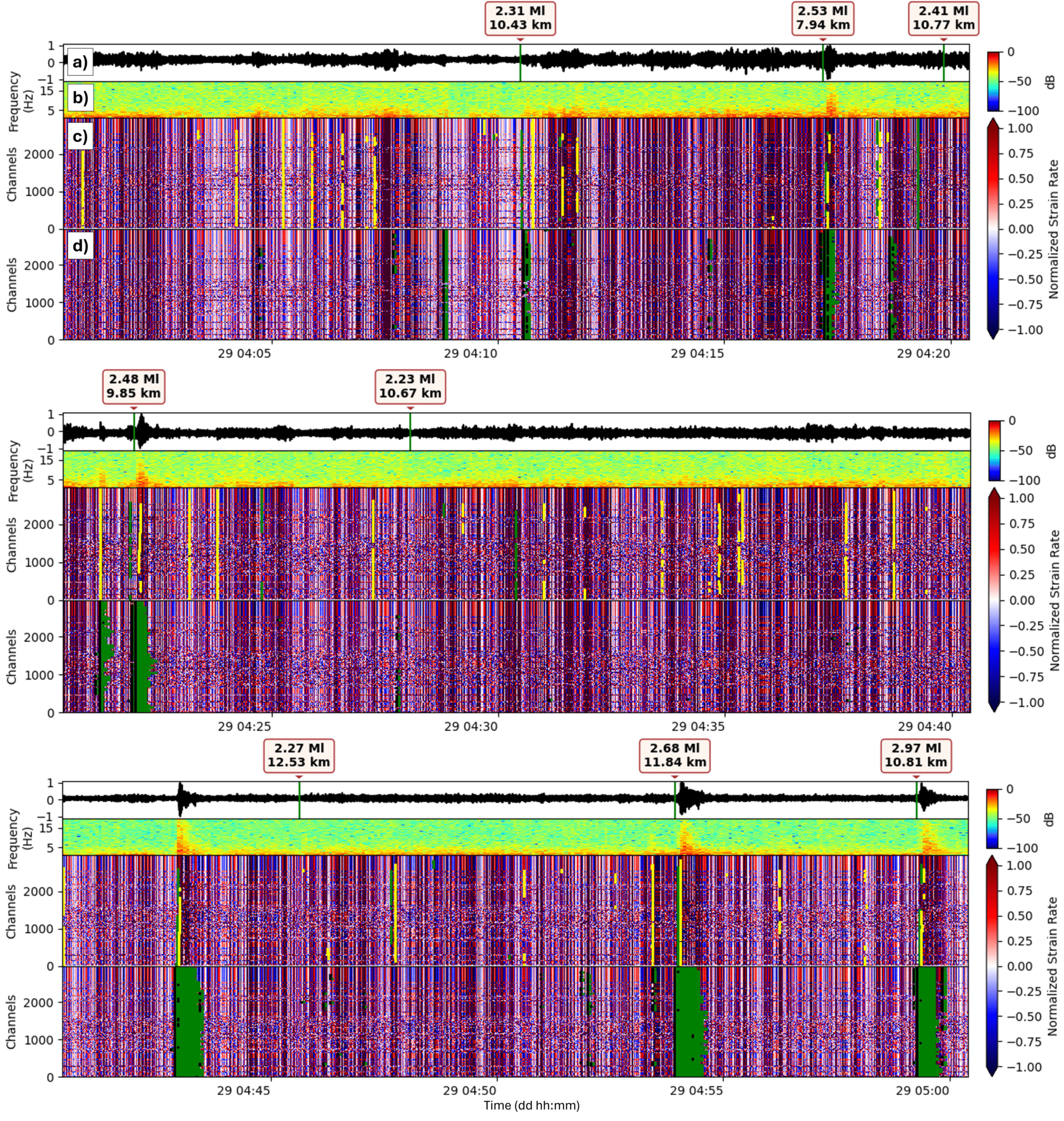}
    \caption{Prediction results for continuous monitoring during the 04:00-05:00 hour on October 29. a) Straingram from channel 1500 with catalog events marked by their origin time, magnitude, and hypocenter. b) Spectrogram from channel 1500. c) Prediction results from the PhaseNet-DAS model, with P phases in green and S phases in yellow. d) Results from the RNN-DAS model, with event start times marked in black and event duration windows in green.}
    \label{fig:webicode}
\end{figure}

First, a generalization experiment of the model was conducted using data from October 29, during the eruption, as this day provides a complete and uninterrupted datastream. The results of applying the RNN-DAS and PhaseNet-DAS models for continuous monitoring over a one-hour period on this day are shown in Figure \ref{fig:webicode}. Both models used the same probability threshold to identify phases or events in each channel.

The results indicate that while the RNN-DAS model successfully matches most events recorded in the DAS catalog that are not attenuated or include regional events absent from the catalog, the PhaseNet-DAS model, although capable of correctly detecting visible events with P and S phases, produces a significant number of false positives in the prediction of S phases. This leads to an overestimation of the number of detected events.

\begin{figure}[H]
    \centering
    \includegraphics[width=\linewidth]{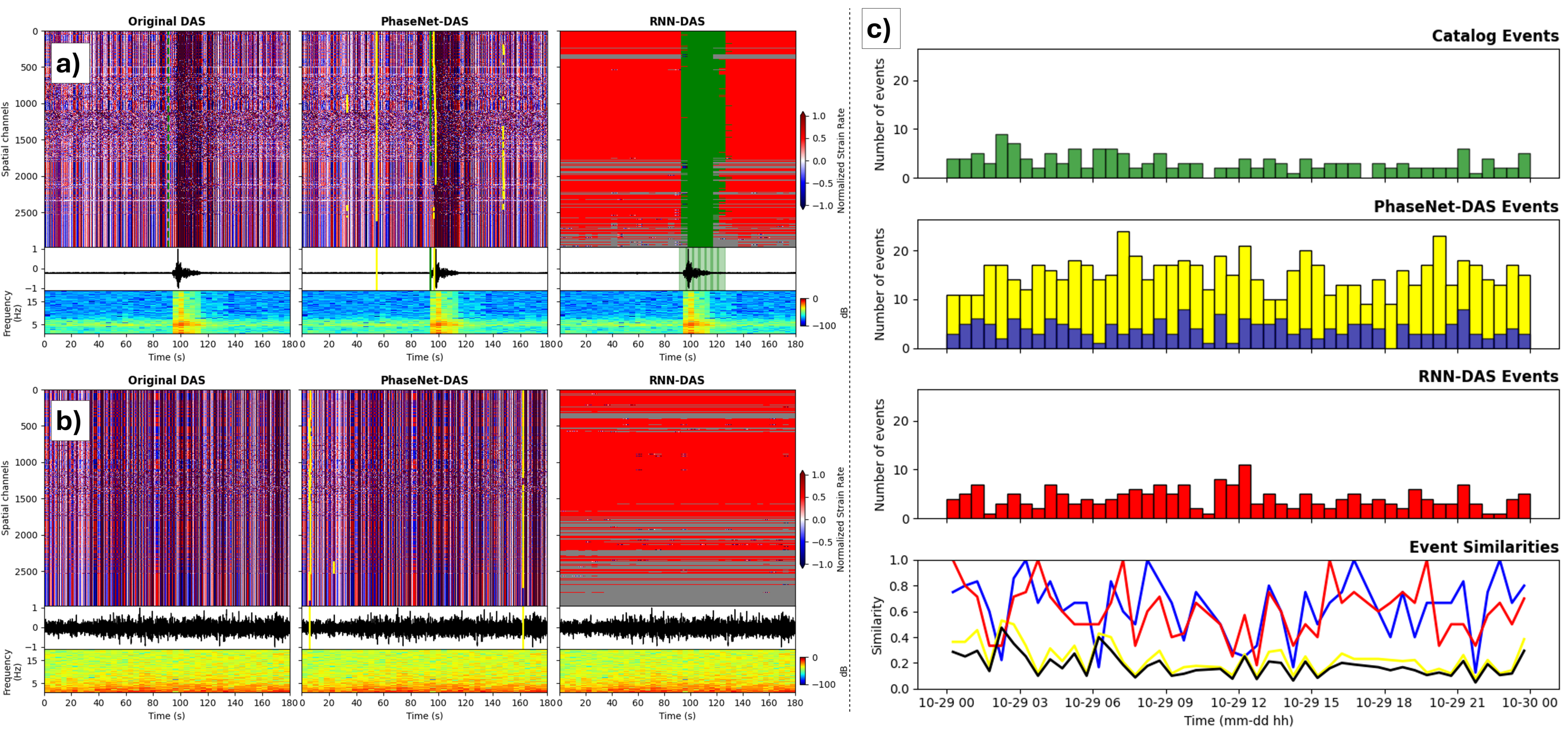}
    \caption{Comparison of PhaseNet-DAS and RNN-DAS model predictions over the 24-hour tested period. a) Event with magnitude $\text{M}_{\text{l}}=3.05$. b) Sample containing only seismic noise. c) Bar plot of catalog events (green), events identified by PhaseNet-DAS (P in blue, S in yellow, and black for joint prediction), and RNN-DAS (red) in 1-hour intervals, along with similarity indices with the catalog.
}
    \label{fig:bars}
\end{figure}

This is more clearly illustrated in Figure \ref{fig:bars} a) and \ref{fig:bars} b), where the overestimation of false positives in the S phase is evident when compared to the RNN-DAS model. Despite this, both models exhibit similar performance when only P-phase picks are considered for event detection in PhaseNet-DAS. In Figure \ref{fig:bars} c), we compare the number of detected events between the two models and the seismic catalog. Using the Jaccard similarity index \citep{fletcher2018}, a higher agreement is observed between events detected via P-phase picks in PhaseNet-DAS and RNN-DAS, though overestimation occurs when S-phase picks are included.

This behavior may be attributed to the low SNR of DAS records due to the volcanic medium through which the waves propagate. In such environments, attenuation and path deviation effects are more significant compared to non-volcanic settings for which the model was originally designed. As a result, the CNN-based architecture of PhaseNet-DAS may encounter greater challenges in accurately detecting seismic phases in those waveforms. In contrast, the RNN-based architecture of RNN-DAS, which incorporates parameterized energy features, appears better suited to distinguishing seismic events from noise in these conditions.

These results showcase that the RNN-DAS model, despite being applied to different time periods than those used for training, where the existing seismicity may differ due to the evolution of the volcanic system, still performs well. This highlights its potential as a volcano monitoring tool over extended periods, during which both the volcanic system and seismicity may evolve, making the model a valuable tool for forecasting.

\begin{figure}[H]
    \centering
    \includegraphics[width=\linewidth]{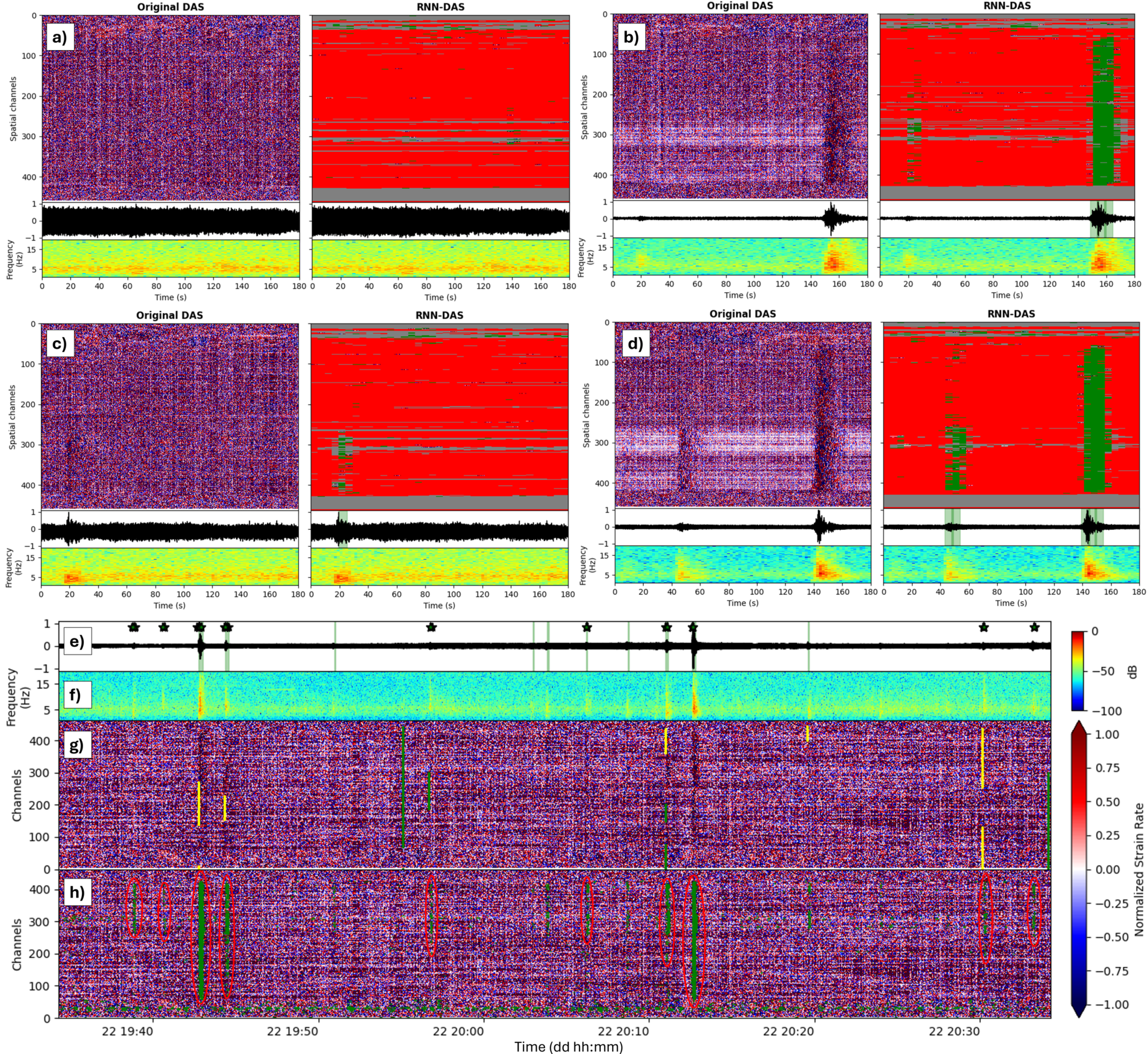}
    \caption{Performance of the RNN-DAS model on DAS data recorded at Stromboli. a) Displays the prediction indicating the absence of events, with only volcanic tremor present. b) and d) illustrate the prediction for events with high SNR. c) Shows the prediction for an event that is masked by volcanic tremor. g) and h) present the continuous predictions over the entire DAS record for the PhaseNet-DAS (with P-phases in green and S-phases in yellow) and RNN-DAS (highlighted in red ovals where an event is detected), respectively. The displayed channel in e) corresponds to channel 400, with its spectrogram shown in f), where the event predictions for this channel are highlighted in green, and a green star marks each window in which RNN-DAS detects an event across all DAS channels.}
    \label{fig:stromboli}
\end{figure}

For the second generalization experiment, no grammar function was employed in order to assess the raw capabilities of the model in this new volcanic setting. The prediction results are shown in Figure \ref{fig:stromboli}. It was found that RNN-DAS is capable of satisfactorily detecting the presence of events with high SNR (Figure \ref{fig:stromboli} b) and d)) as well as their absence in records containing only volcanic tremor (Figure \ref{fig:stromboli} a)). In the case of smaller events with low SNR (Figure \ref{fig:stromboli} c)), the model still identifies the events; however, due to the high-amplitude volcanic tremor, the events are masked, which hinders the model's ability to detect them across all channels (Figure \ref{fig:stromboli} c)). Nonetheless, the results for detecting explosive events (Figiure \ref{fig:stromboli} e) and f)) are similar to those reported by \cite{biagioli2023}. For instance, the PhaseNet-DAS model does not accurately recognize the presence of these events (Figure \ref{fig:stromboli} g)), due to the clear absence of seismic phases and the significant high-amplitude tremor that masks the waveforms.

\begin{figure}[H]
    \centering
    \includegraphics[width=\linewidth]{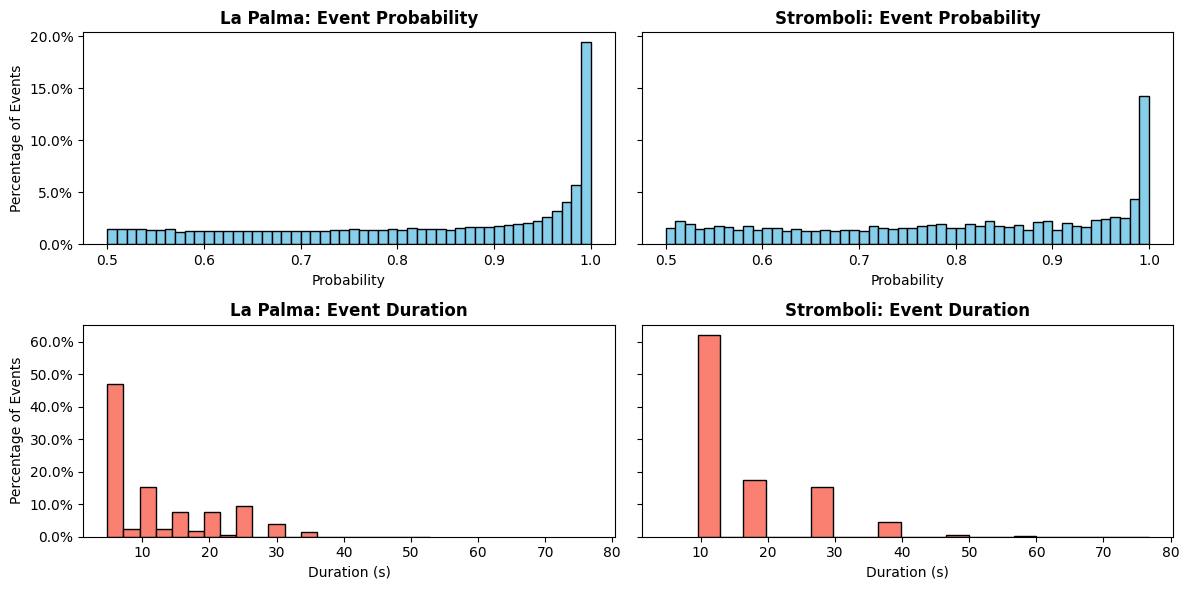}
    \caption{Bar plots of the mean detection probability and duration for La Palma (left) and Stromboli (right), based on detections per DAS channel in the RNN-DAS model generalization tests. The y-axis values represent the percentage of detections corresponding to each interval.}
    \label{fig:bars_lapalma_stromboli}
\end{figure}

It is important to note that, while RNN-DAS can detect seismic events associated with volcanic explosions, it cannot distinguish between seismic noise and volcanic tremor, nor can it differentiate volcanic explosions from purely tectonic (VT) events. This limitation arises because the model has not been trained to differentiate between these types of events. However, this issue could easily be addressed by retraining the model and redefining the prediction classes to better adapt to the characteristic events of each volcano. For instance, when comparing the results from both generalization tests for RNN-DAS predictions (see Figure \ref{fig:bars_lapalma_stromboli}), it is observe that, while event durations show a similar distribution for both regions, there is a slight difference in detection probabilities. Although small, this difference can be attributed to the distinct source mechanisms of the events despite their similarities in energy or frequency range, with VT events from La Palma having a higher detection probability compared to volcanic explosions from Stromboli. Thus suggesting that the model’s ability to distinguish between these events could be further enhanced through retraining, enabling it to better differentiate between various volcano-seismic event types. This refinement would improve the model's performance and adaptability, allowing it to more accurately identify and classify different events detected by DAS systems in active volcanic regions.

\section{Conclusions}
In this paper, we have demonstrated how RNNs with LSTM can be applied to the study of seismovolcanic signals for volcanic monitoring purposes using the DAS technique. To achieve this, we developed a new deep learning model, RNN-DAS, designed to operate in real-time for detecting volcanic-tectonic events and recording their waveforms. The model leverages features extracted from the parametrization of these signals into energy coefficients, which account for both the temporal evolution and spatial coherence provided by ultra-dense DAS measurements.

The results show that the model achieves over 90\% in all studied metrics, yielding performance comparable to visual inspection and surpassing other existing models. Trained with a large dataset of VT events from an underwater high-fidelity DAS during the La Palma 2021 eruption, the model has proven capable of detecting volcanic events in generalization tests across different time periods. It has demonstrated reliability for the same volcano, as well as adaptability for others, such as Stromboli. Despite changes in source mechanisms, medium effects, or in the case of land-based DAS, the model continues to provide accurate predictions on volcano-seismic events recorded by DAS arrays.

This advancement addresses key challenges in volcanic seismology, particularly in managing the vast datasets produced by DAS, and represents a significant step forward by enabling automated, scalable, and real-time analysis of volcanic signals. Looking ahead, two promising lines of research are proposed to further enhance the model's adaptability. First, testing its real-time applicability with other DAS datasets of VT events from active volcanoes will broaden its monitoring capabilities. Second, retraining the model with new DAS data from different types of volcano-seismic events, such as those observed at Stromboli, will ensure more comprehensive and accurate classification in diverse volcanic environments by redefining classification classes to better accommodate the different types of events detected.

\section*{Appendix A: RNN Architecture and LSTM Structure}
This appendix provides a more detailed explanation of the architecture of Recurrent Neural Networks (RNNs) with Long Short-Term Memory (LSTM) cells used in the RNN-DAS model.

\subsection*{Recurrent Neural Networks (RNNs)}
A RNN can be understood as a type of artificial neural network specifically designed to process sequential data. Unlike classical feedforward models, an RNN maps an input $ x_t $ to an output $ y_t $ as follows \citep{kamath2019}:
\begin{equation}
    \text{RNN}(x_t, h_{t-1})\mapsto (y_t, h_t) \equiv y_t = \text{softmax}\left( f_{\theta}(x_t, h_{t-1}) \right),
\end{equation}
where $ h_t $ represents the hidden state at time $ t $ which acts as the fundamental memory unit containing information up to and including the time step $t$. After applying the softmax function to normalize per class-probabilities, we obtain the final output at $t$. The RNN captures temporal dependencies across inputs through the operation of unfolding (see Figure \ref{fig:arquitectura} (a)), where the hidden state is fed back into the network at each time step. This process expands the recurrent structure over multiple time steps, allowing the output at time $ t $ to depend not only on the current input but also on past inputs through the hidden state via a recurrent process. These nonlinear operations enable the RNN to effectively model sequential data by maintaining memory of prior states.

Additionally, the model parameters $ \theta $ are shared across all time steps in weight matrices, effectively treating each step as a layer in a globally deep forward neural network \citep{lecun2015}.

This mapping $ f_{\theta} $ can be formally expressed using the weight matrices $ U $ (input to hidden connections), $ W $ (hidden to hidden), and $ V $ (hidden to output), which act on the input vector $ x_t $ as follows:
\begin{equation}
f_{\theta}(x_t, h_{t-1}) = \left( V \cdot \sigma(x_t \cdot U + h_{t-1} \cdot W + b) \right) = V \cdot h_t,
\end{equation}
where $ b $ is the bias vector, and $ \sigma $ represents a nonlinear activation function, usually the sigmoid function. By maintaining the matrix $ W $ shared across all unfolding operations, the weights are shared, thereby achieving a memory sensitive to temporal variations \citep{titos2018b}. This is true for the case of only one layer of RNN, but just as with fully connected ANNs or CNNs, several layers of RNNs can be combined, resulting in deep RNNs \citep{kamath2019, pascanu2013a, ElHihi1996}.

Another key difference between feed-forward ANNs and RNNs, with respect to the training process, is that RNNs do not rely solely on the current time step. This means that, during training, the traditional backpropagation with stochastic gradient descent is not used. Instead, errors must be propagated across all time steps, a process known as backpropagation through time (BPTT) \citep{schmidhuber2015}. This introduces the challenge of what is known as \textit{vanishing} or \textit{exploding} gradients \citep{pascanu2013a}, which significantly limits the ability of RNNs to model temporal dependencies. Specifically, if the gradients of the loss function become too small or too large for any given time step, they may propagate through time and exacerbate the problem exponentially \citep{Bengio1994}. 
To address this issue, several techniques have been proposed, including careful tuning of hyperparameters or the use of adaptive learning rates \citep{Kingma2014}. However, the most widely used method involves adding gates to the flow of information in memory, giving rise to models such as Long Short-Term Memory (LSTM).

\subsection*{Long Short-Term Memory (LSTM)}
\begin{figure}[H]
    \centering
    \includegraphics[width=1\linewidth]{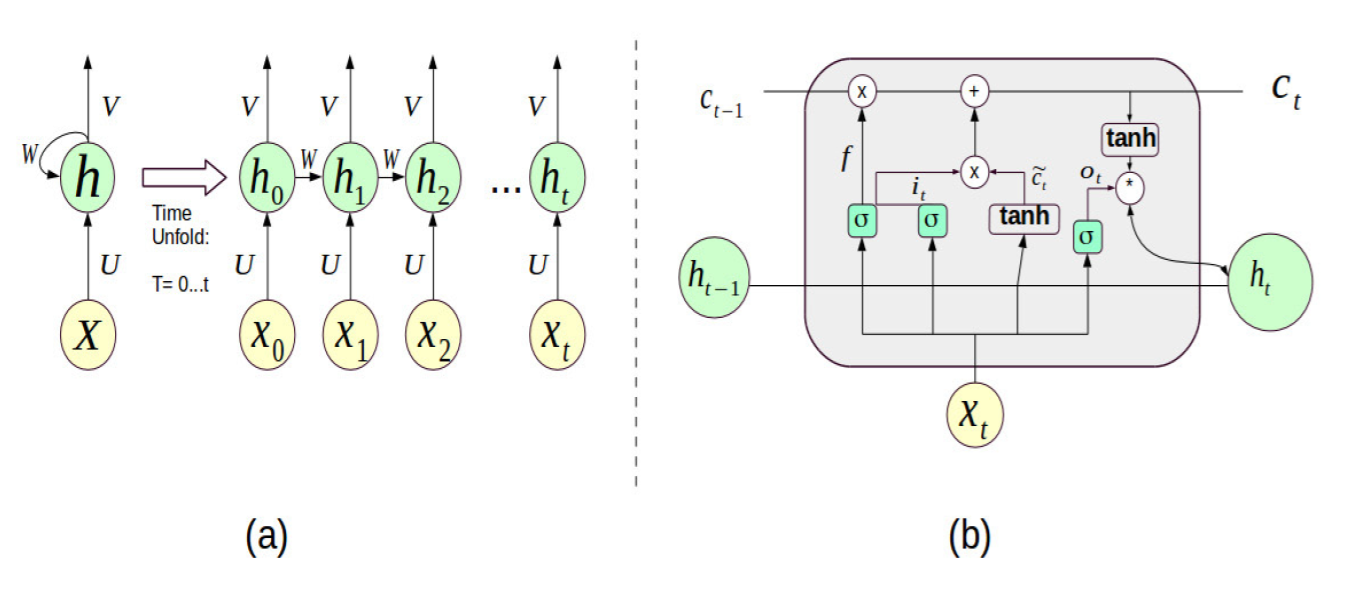}
    \caption{(a) Operation of time unfolding at time step $ t $ in a vanilla RNN, showing the relationship between inputs, hidden states, and output states associated with the matrices $ U $, $ V $, and $ W $.  
    (b) LSTM cell with input $ i $, output $ o $, forget gates $ f $, and memory candidates $ c $ at time step $ t $.  Modified from \cite{titos2018b}.}
    \label{fig:arquitectura}
\end{figure}

Long Short-Term Memory (LSTM) networks use specialized gates to regulate the gradient flow within the memory of the recurrent neural network \citep{Hochreiter1997}. These gates, namely the input ($ i_t $), output ($ o_t $), and forget gate ($ f_t $), are used to control the memory cell and have their respective $ U $ and $ W $ weight matrices. This memory cell (see Figure~\ref{fig:arquitectura} (b)) passes the hidden state to the next time step with the new modified memory content, as the gates themselves are implemented as neural network layers, allowing the model to learn when to store, discard, or update information in the memory cell.

This is done via the computation of a series of memory candidates $ \bar{c}_t $ for each time step, from which the memory content $ c_t $ is selected after applying these gates as follows:
\begin{align}
    & i_t = \sigma(W^i \cdot x_t + U^i \cdot h_{t-1} + b^i), \\
    & f_t = \sigma(W^f \cdot x_t + U^f \cdot h_{t-1} + b^f), \\
    & o_t = \sigma(W^o \cdot x_t + U^o \cdot h_{t-1} + b^o), \\
    & \bar{c}_t = \text{tanh}(W^c \cdot x_t + U^c \cdot h_{t-1}) \Rightarrow c_t = f_t \cdot c_{t-1} + i_t \cdot \bar{c}_t.
\end{align}
Once the memory content at time $ t $, which depends on the content at $ t-1 $, is obtained, the hidden state $ h_t $ is computed as:
\begin{equation}
    h_t = o_t \cdot \text{tanh}(c_t),
\end{equation}
and to obtain the output probabilities per class at time $ t $, it is sufficient to apply the output matrix of weights $V$ and normalize using softmax as in equation (1).

\section*{Data Availability}

The datasets, the RNN-DAS model, and the necessary codes for its implementation are available upon reasonable request to the corresponding author, Javier Fernández-Carabantes (\href{mailto:javierfyc@ugr.es}{javierfyc@ugr.es}).

\section*{Acknowledgments}

This work was developed as part of the DigiVolCan project – A digital infrastructure for forecasting volcanic eruptions in the Canary Islands. 
The project was funded by the Ministry of Science, Innovation, and Universities / State Research Agency (MICIU/AEI) of Spain, and the European Union through the Recovery, Transformation, and Resilience Plan, Next Generation EU Funds. Project: PLEC2022-009271 funded by MICIU/AEI /10.13039/501100011033 and by the European Union Next GenerationEU/ PRTR.

\bibliographystyle{elsarticle-harv} 
{\footnotesize \bibliography{Refs}}

\end{document}